\RequirePackage{lineno}
\documentclass[floatfix,twocolumn,showpacs,preprintnumbers,amsmath,amssymb,prl,superscriptaddress]{revtex4} 
\usepackage{url}
\usepackage{graphicx}
\usepackage{dcolumn}
\usepackage{bm}
\usepackage{soul}

\usepackage{color}
\usepackage{natbib}
\begin{document}
\raggedbottom
\title{
\begin{flushright}
{\small  STAR Review: version A1~
\today \\
 }
\end{flushright} 
First observation of the directed flow of $D^{0}$ and $\overline{D^0}$
in Au+Au collisions at $\sqrt{s_{\rm NN}}$ = 200\,GeV}

\affiliation{Abilene Christian University, Abilene, Texas   79699}
\affiliation{AGH University of Science and Technology, FPACS, Cracow 30-059, Poland}
\affiliation{Alikhanov Institute for Theoretical and Experimental Physics, Moscow 117218, Russia}
\affiliation{Argonne National Laboratory, Argonne, Illinois 60439}
\affiliation{Brookhaven National Laboratory, Upton, New York 11973}
\affiliation{University of California, Berkeley, California 94720}
\affiliation{University of California, Davis, California 95616}
\affiliation{University of California, Los Angeles, California 90095}
\affiliation{University of California, Riverside, California 92521}
\affiliation{Central China Normal University, Wuhan, Hubei 430079 }
\affiliation{University of Illinois at Chicago, Chicago, Illinois 60607}
\affiliation{Creighton University, Omaha, Nebraska 68178}
\affiliation{Czech Technical University in Prague, FNSPE, Prague 115 19, Czech Republic}
\affiliation{Technische Universit\"at Darmstadt, Darmstadt 64289, Germany}
\affiliation{E\"otv\"os Lor\'and University, Budapest, Hungary H-1117}
\affiliation{Frankfurt Institute for Advanced Studies FIAS, Frankfurt 60438, Germany}
\affiliation{Fudan University, Shanghai, 200433 }
\affiliation{University of Heidelberg, Heidelberg 69120, Germany }
\affiliation{University of Houston, Houston, Texas 77204}
\affiliation{Huzhou University, Huzhou, Zhejiang  313000}
\affiliation{Indiana University, Bloomington, Indiana 47408}
\affiliation{Institute of Physics, Bhubaneswar 751005, India}
\affiliation{University of Jammu, Jammu 180001, India}
\affiliation{Joint Institute for Nuclear Research, Dubna 141 980, Russia}
\affiliation{Kent State University, Kent, Ohio 44242}
\affiliation{University of Kentucky, Lexington, Kentucky 40506-0055}
\affiliation{Lawrence Berkeley National Laboratory, Berkeley, California 94720}
\affiliation{Lehigh University, Bethlehem, Pennsylvania 18015}
\affiliation{Max-Planck-Institut f\"ur Physik, Munich 80805, Germany}
\affiliation{Michigan State University, East Lansing, Michigan 48824}
\affiliation{National Research Nuclear University MEPhI, Moscow 115409, Russia}
\affiliation{National Institute of Science Education and Research, HBNI, Jatni 752050, India}
\affiliation{National Cheng Kung University, Tainan 70101 }
\affiliation{Nuclear Physics Institute of the CAS, Rez 250 68, Czech Republic}
\affiliation{Ohio State University, Columbus, Ohio 43210}
\affiliation{Institute of Nuclear Physics PAN, Cracow 31-342, Poland}
\affiliation{Panjab University, Chandigarh 160014, India}
\affiliation{Pennsylvania State University, University Park, Pennsylvania 16802}
\affiliation{NRC "Kurchatov Institute", Institute of High Energy Physics, Protvino 142281, Russia}
\affiliation{Purdue University, West Lafayette, Indiana 47907}
\affiliation{Pusan National University, Pusan 46241, Korea}
\affiliation{Rice University, Houston, Texas 77251}
\affiliation{Rutgers University, Piscataway, New Jersey 08854}
\affiliation{Universidade de S\~ao Paulo, S\~ao Paulo, Brazil 05314-970}
\affiliation{University of Science and Technology of China, Hefei, Anhui 230026}
\affiliation{Shandong University, Qingdao, Shandong 266237}
\affiliation{Shanghai Institute of Applied Physics, Chinese Academy of Sciences, Shanghai 201800}
\affiliation{Southern Connecticut State University, New Haven, Connecticut 06515}
\affiliation{State University of New York, Stony Brook, New York 11794}
\affiliation{Temple University, Philadelphia, Pennsylvania 19122}
\affiliation{Texas A\&M University, College Station, Texas 77843}
\affiliation{University of Texas, Austin, Texas 78712}
\affiliation{Tsinghua University, Beijing 100084}
\affiliation{University of Tsukuba, Tsukuba, Ibaraki 305-8571, Japan}
\affiliation{United States Naval Academy, Annapolis, Maryland 21402}
\affiliation{Valparaiso University, Valparaiso, Indiana 46383}
\affiliation{Variable Energy Cyclotron Centre, Kolkata 700064, India}
\affiliation{Warsaw University of Technology, Warsaw 00-661, Poland}
\affiliation{Wayne State University, Detroit, Michigan 48201}
\affiliation{Yale University, New Haven, Connecticut 06520}

\author{J.~Adam}\affiliation{Creighton University, Omaha, Nebraska 68178}
\author{L.~Adamczyk}\affiliation{AGH University of Science and Technology, FPACS, Cracow 30-059, Poland}
\author{J.~R.~Adams}\affiliation{Ohio State University, Columbus, Ohio 43210}
\author{J.~K.~Adkins}\affiliation{University of Kentucky, Lexington, Kentucky 40506-0055}
\author{G.~Agakishiev}\affiliation{Joint Institute for Nuclear Research, Dubna 141 980, Russia}
\author{M.~M.~Aggarwal}\affiliation{Panjab University, Chandigarh 160014, India}
\author{Z.~Ahammed}\affiliation{Variable Energy Cyclotron Centre, Kolkata 700064, India}
\author{I.~Alekseev}\affiliation{Alikhanov Institute for Theoretical and Experimental Physics, Moscow 117218, Russia}\affiliation{National Research Nuclear University MEPhI, Moscow 115409, Russia}
\author{D.~M.~Anderson}\affiliation{Texas A\&M University, College Station, Texas 77843}
\author{R.~Aoyama}\affiliation{University of Tsukuba, Tsukuba, Ibaraki 305-8571, Japan}
% ...

\collaboration{STAR Collaboration}\noaffiliation

\begin{abstract}
We report the first measurement of rapidity-odd directed flow ($v_{1}$) for $D^{0}$ and $\overline{D^{0}}$ mesons at mid-rapidity ($\lvert y \rvert \textless 0.8$) in Au+Au collisions at $\sqrt{s_{\rm NN}}$ = 200\,GeV using the STAR detector at the Relativistic Heavy Ion Collider. In 10--80\% Au+Au collisions, the slope of the $v_{1}$ rapidity dependence ($dv_{1}/dy$), averaged over $D^{0}$ and $\overline{D^{0}}$ mesons, is -0.080 $\pm$ 0.017 (stat.) $\pm$ 0.016 (syst.) for transverse momentum $p_{\rm T}$ above 1.5~GeV/$c$. The absolute value of $D^0$-meson $dv_1/dy$ is about 25 times larger than that for charged kaons, with 3.4$\sigma$ significance. These data give a unique insight into the initial tilt of the produced matter, and offer constraints on the geometric and transport parameters of the hot QCD medium created in relativistic heavy-ion collisions.
\end{abstract}

\maketitle
An important goal of relativistic heavy-ion collisions is to understand the production and dynamics of strongly interacting matter produced at high energy densities~\cite{qgp1,qgp2,qgp3,qgp4,whitepaper1,whitepaper2,whitepaper3,whitepaper4}. The collective motion of particles emitted in such collisions are of special interest because of their sensitivity to the initial stages of the collision, when production of a deconfined Quark-Gluon Plasma (QGP) phase is expected. The directed flow ($v_{1}$) of particles is characterized by the first harmonic Fourier coefficient in the azimuthal distribution relative to the reaction plane~\cite{methods1, methods2, methods3}. A hydrodynamic calculation with a tilted initial QGP source~\cite{tilt_chv1_bozek} can explain the observed negative $v_{1}$ slope or ``anti-flow''~\cite{v1_anti} near midrapidity, for charged hadrons measured at RHIC energies~\cite{v1-62, v1-4systems, pidv1_200}. However, additional contributions to the directed flow could result from a dipole-like density asymmetry, nuclear shadowing (the interactions between particles and spectators), or a difference in density gradients in different directions within the transverse plane~\cite{raimond_transport, heinz, cuau_vn}. The study of heavy quarks ($c$ and $b$) in heavy-ion collisions is especially important due to their early creation. Owing to their large masses, heavy quarks are predominantly produced in initial hard scatterings and their relaxation time in the QGP medium is comparable to the lifetime of the QGP. Consequently, heavy quarks are an excellent probe to study QGP dynamics~\cite{hf_review}. 

The transverse momentum ($p_{\rm T}$) spectra and elliptic flow ($v_{2}$) of $D^{0}$ mesons at midrapidity have been measured at RHIC~\cite{star_d0_v2_ncq,star_d0_raa_err} and LHC~\cite{ALICE-HFv2_1, ALICE-HFv2_2,ALICE-HF-D-RAA} energies. The magnitude of $v_{2}$ for the charm hadrons is found to follow the number-of-constituent-quark (NCQ) scaling pattern observed for light hadron species in non-central heavy-ion collisions~\cite{ncq_star1,ncq_star2,ncq_phenix1,star_d0_v2_ncq}. Furthermore, charm hadron yields are observed to be significantly suppressed at high $p_{\rm T}$, similar to light hadron species in central heavy-ion collisions. Simultaneous descriptions of charm $v_{2}$ and nuclear modification factors ($R_{\rm AA}$)~\cite{RAA_star,RAA_phenix,star_d0_raa,star_d0_raa_err} have been used to constrain the QGP transport parameters for heavy quarks, such as its drag and diffusion coefficients.

A recent model calculation utilizing Langevin dynamics coupled to a hydrodynamic medium with a tilted initial source, predicted a significantly larger $v_{1}$ for $D$-mesons compared to light flavor hadrons~\cite{chatt_bozek_dv1_prl}. A notable feature is the strong sensitivity of $D$-meson $v_{1}$ to the initial tilt of the QGP source compared to that of light hadrons. The magnitude of the observed heavy quark $v_{1}$ is also sensitive to the QGP transport parameters in the hydrodynamic calculation.

It is further predicted that the transient magnetic field generated in
heavy-ion collisions can induce a larger directed flow for heavy quarks than for light quarks due to the Lorentz
force~\cite{das_greco_dv1_em, gursoy_em}. The $v_1$ induced by this
initial electromagnetic (EM) field is expected to have the same
magnitude, but opposite charge sign for charm ($c$) and anti-charm
($\bar{c}$) quarks. This suggests that the $v_{1}$ measurements of heavy quarks could offer crucial insight into the properties of the initial EM field. A hydrodynamic model calculation which includes both the initially tilted source and the EM field predicts that the $D$-mesons will have a significant $v_1$ as a function of rapidity ($y$) and a splitting is to be expected between $D$-mesons and $\overline{D}$-mesons due to the initial magnetic field~\cite{chatt_bozek_dv1_2nd}.

In this Letter, we report the first measurement of rapidity-odd
directed flow for $D^{0}$ and $\overline{D^{0}}$ mesons in Au+Au
collisions at $\sqrt{s_{\rm NN}}$ = 200\,GeV in the STAR
experiment~\cite{starnim}. We utilize the Heavy Flavor Tracker
(HFT)~\cite{hft_tdr, hft_tdr2}, a high-resolution silicon detector
consisting of four cylindrical layers. Beginning at the largest
radius, there is one layer of Silicon Strip Detector (SSD), one layer
of Intermediate Silicon Tracker (IST), and two layers of Pixel
Detectors (PXL). The reconstruction of heavy-flavor hadrons is greatly
enhanced due to the excellent track pointing resolution and secondary
vertex resolution offered by the HFT. STAR collected minimum-bias (MB)
triggered events with the HFT during the years 2014 and 2016. 
The MB events were selected by a coincidence between the
east and west Vertex Position Detectors (VPD)~\cite{VPD} located at
pseudorapidity $4.4 < |\eta| < 4.9$. To ensure good HFT acceptance,
the reconstructed primary vertex along the $z$-direction is required
to be within 6 cm of the center of the detector. Approximately 2.2
billion MB triggered good quality events are used in this analysis.

The $D^{0}$ and $\overline{D^{0}}$ mesons are reconstructed via their hadronic decay channel: $D^{0} (\overline{D^{0}}) \rightarrow K^{-} \pi^{+} (K^{+} \pi^{-})$ (branching fraction 3.89\%, $c \tau \sim 123  \;  \mu$m). Hereafter, $D^0$ refers to the combined $D^{0}$ and $\overline{D^{0}}$ samples, unless 
explicitly stated otherwise. The charged particle tracks are
reconstructed using the Time Projection Chamber (TPC)~\cite{startpc}
together with the HFT in a uniform 0.5 T magnetic field. The collision
centrality is determined from the number of charged particles within
$|\eta| < 0.5$ and corrected for trigger inefficiency using a Monte
Carlo Glauber simulation~\cite{centrality_glauber}. Good quality tracks are ensured
by requiring a minimum of 20 TPC hits (out of a possible 45), hits in
both layers of PXL, at least one hit in the IST or SSD layer. Further,
the tracks are required to have transverse momentum $p_{\rm T} > 0.6$
GeV/$c$ and pseudorapidity $|\eta| < 1$. The $D^{0}$ decay daughters
are identified via specific ionization energy loss ($dE/dx$)
inside the TPC and from $1/\beta$ measurements by the Time of Flight
(TOF)~\cite{TOF} detector. To identify particle species, the $dE/dx$
is required to be within three and two standard deviations from the
expected values for $\pi$ and $K$, respectively. When tracks are
associated with the hits in the TOF detector, the $1/\beta$ is required to be within three standard deviations from the expected values for both $\pi$ and $K$.

The $D^{0}$ decay vertex is reconstructed as the mid-point of 
the distance of closest approach between the two decay daughter
tracks. Background arises due to random combinations of tracks passing close to the collision point. The decay topological cuts are tuned to reduce the background and enhance the signal-to-background ratio. The topological cut variables are optimized using the Toolkit for Multivariate Data Analysis (TMVA) package~\cite{tmva} and are discussed in Refs.~\cite{star_d0_v2_ncq, star_d0_raa}. 

\begin{figure}[!htb]
\centering
\centerline{\includegraphics[scale=0.47]{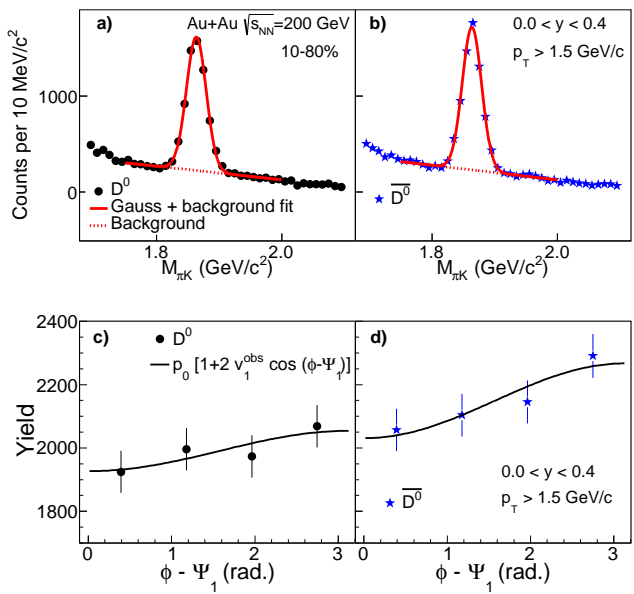}}
\caption{$D^{0}$ (panel (a)) and $\overline{D^{0}}$ (panel (b))
  invariant mass distribution for $0.0 < y < 0.4$ and $p_{\rm T} >$ 1.5~GeV/$c$ in 10--80\% central Au+Au collisions at $\sqrt{s_{\rm NN}}$ = 200~GeV. The solid line represents a Gaussian fit plus a linear
  function for the random combinatorial background.
$D^{0}$ (panel (c)) and $\overline{D^{0}}$ (panel (d)) yields in azimuthal angle bins relative to the first-order event-plane azimuth ($\phi - \Psi_{1}$) for $0.0 < y < 0.4$ and $p_{\rm T} >$ 1.5~GeV/$c$ in 10--80\% central Au+Au collisions at $\sqrt{s_{\rm NN}}$ = 200~GeV. The solid line presents a fit to the function $p_{0}[1 + 2 v_{1}^{\rm obs}\cos (\phi - \Psi_{1})]$. Vertical bars show statistical uncertainties.}
\label{fig1}
\end{figure}
The first-order event plane ($\Psi_{1}$) is measured by using the east
and west Zero Degree Calorimeter Shower Maximum Detectors
(ZDC-SMD)~\cite{zdc-thesis,v1-62,v1-4systems,pidv1_200, STAR-BESv1},
which are located at $|\eta| > 6.3$. Since the $v_{1}$ signal is
strong at forward rapidity, the ZDC-SMD provides better first-order
event plane resolution than detectors closer to midrapidity. Moreover,
the five units of $\eta$ gap between the ZDC-SMDs and the TPC and HFT
significantly reduce possible systematic error in $v_1$ arising from
non-flow effects \cite{methods2, methods3}. Such effects could result
from resonances, jets, quantum statistics, and final-state
interactions like Coulomb effects. Systematic uncertainties arising
from event-plane estimation are at the level of less than 2\% and are discussed in Ref.~\cite{STAR-BESv1}.

The $D^{0}$ $v_{1}$ is calculated using the event plane
method~\cite{methods1, methods2, methods3}. Figures~\ref{fig1}(a) and
\ref{fig1}(b) show the $D^{0}$ and $\overline{D^{0}}$ invariant mass
spectra for $0.0 < y < 0.4$ and $p_{\rm T} >$ 1.5~GeV/$c$ in 10--80\%
central Au+Au collisions at $\sqrt{s_{\rm NN}}$ = 200\,GeV. The choice
of 10--80\% centrality is driven by the fact that the first-order
event plane resolution from ZDC-SMD drops considerably in the 0-10\%
central collisions. The $D^0$ acceptance, in rapidity and azimuthal
angle, under such kinematic selection cuts is uniform across the
measured rapidity region. The invariant mass distributions were fitted
with a Gaussian plus a first-order linear polynomial function. The
linear function provides a good estimate of the random combinatorial
background. The yield is obtained by integrating the distribution in
the range 1.82$-$1.91\,GeV/$c^{2}$ and subtracting the
background beneath the signal. The $D^{0} (\overline{D^{0}})$ yield is
obtained in each $\phi - \Psi_{1}$ bin in four rapidity
windows. Figures~\ref{fig1}(c)  and ~\ref{fig1}(d) present $D^{0}$ and
$\overline{D^{0}}$ yields as a function of $\phi- \Psi_{1}$ for
$0.0<y<0.4$. The value of $v_{1}$ is calculated by fitting the data
with a functional form $p_{0} [1 + 2 v_{1}^{\rm obs} \cos (\phi -
\Psi_{1})]$, indicated by the solid lines in the figure. The ZDC-SMD
event plane resolution correction factors are obtained in seven
centrality bins. For a wide centrality bin (10--80\%), it is
determined from the $D^{0}$-yield-weighted mean of the individual
centrality bins' resolutions using a procedure detailed in
Ref.~\cite{ep_resol}. The final $v_{1}$ is corrected by scaling
$v_1^{\rm obs}$ with the event plane resolution (0.363).

Systematic uncertainties are assessed by comparing the $v_{1}$
obtained from various methods. These comparisons include (i) the fit
vs. side-band methods for the background estimation and (ii) various
invariant mass fitting ranges and residual background functions
(first-order vs. second-order polynomials) for signal extractions,
(iii) histogram bin counting vs. functional integration for yield
extraction, (iv) varying topological cuts so that the efficiency
changes by $\pm$ 50\% with respect to the nominal value, (v) varying
event and track level quality cuts (vi) varying particle
identification cuts. The above comparisons are varied independently to form multiple combinations. For the final systematic uncertainty on the $v_{1}(y)$ and $dv_{1}/dy$, the difference between the default settings and alternative measurements from these sources are added in quadrature. Further, the systematic uncertainty in each rapidity bin is symmetrized by considering the maximum uncertainty between $D^{0}$ and $\overline{D^{0}}$.
\begin{figure}[!htb]
\begin{center}
\includegraphics[scale=0.45]{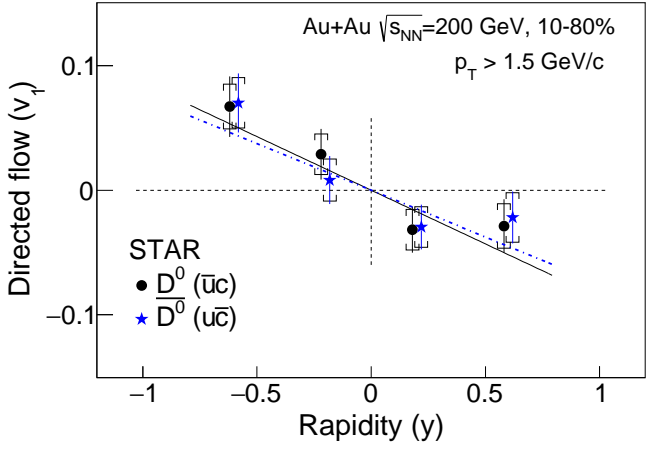}
\caption{Filled circles and star symbols present $v_{1}$ as a function of rapidity for $D^{0}$ and $\overline{D^{0}}$ mesons at $p_{\rm T} > $1.5~GeV/$c$ for 10--80\% centrality Au+Au collisions at $\sqrt{s_{\rm NN}}$ = 200 GeV. The $D^{0}$ and $\overline{D^{0}}$ data points are displaced along the $x$-axis by $\mp$ 0.019 respectively for clear visibility. The error bars and caps denote statistical and systematic uncertainties, respectively. The solid and dot-dashed lines present a linear fit 
to the data points for $D^{0}$ and $\overline{D^{0}}$, respectively.}
\label{fig2}
\end{center}
\end{figure}

\begin{figure}[!htb]
\begin{center}
\includegraphics[scale=0.45]{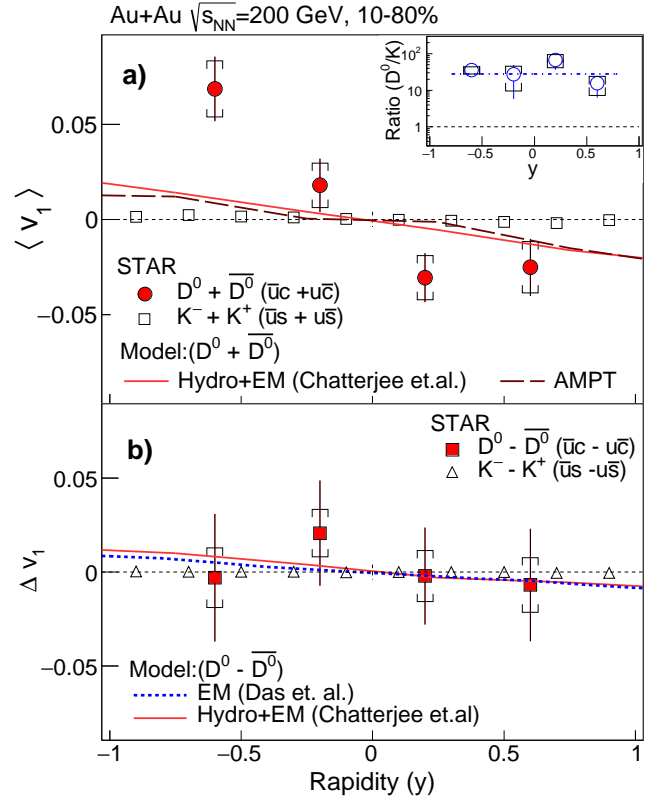}
\caption{ Panel (a): Solid circles present directed flow ($\langle
  v_1(y) \rangle$) for the combined samples of $D^{0}$ and
  $\overline{D^{0}}$ at $p_{\rm T} > 1.5$~GeV/$c$ in 10--80\%
  central Au+Au collisions at $\sqrt{s_{\rm NN}}$ = 200\,GeV. Open
  squares present $v_1(y)$ for charged kaons with $p_{\rm T} > $0.2
  GeV/$c$. The inset shows the ratio of $v_{1}$ between the $D^{0}$
  and charged kaons. The solid  and dashed lines show hydrodynamic
  model calculation with an initial electromagnetic
  field~\cite{chatt_bozek_dv1_prl,chatt_bozek_dv1_2nd} and AMPT
  model~\cite{ampt_dv1} calculations, respectively. Panel (b): The
  solid square markers present the difference in $v_1(y)$ ($\Delta
  v_{1}$) between $D^{0}$ and $\overline{D^{0}}$ for $p_{\rm T} > $1.5~GeV/$c$ in 10--80\% Au+Au collisions at $\sqrt{s_{\rm NN}}$ = 200
  GeV. Open triangles represent $\Delta v_{1}$ between $K^{-}$ and
  $K^{+}$. The dotted and solid lines present a $\Delta v_{1}$
  prediction for $D^{0}$ and $\overline{D^{0}}$,  reported in
  Refs.~\cite{das_greco_dv1_em} and
  ~\cite{chatt_bozek_dv1_prl,chatt_bozek_dv1_2nd}, respectively. The error bars and caps denote statistical and systematic uncertainties, respectively.}  
\label{fig3}
\end{center}
\end{figure}

In Fig.~\ref{fig2}, the filled circle and star markers present the
rapidity dependence of $v_{1}$ for the $D^{0}$ and $\overline{D^{0}}$
mesons with $p_{\rm T} >$ 1.5~GeV/c in 10--80$\%$ Au+Au collisions at
$\sqrt{s_{\rm NN}}$ = 200\,GeV. It is a common practice to present the
strength of $v_{1}$ via its slope at midrapidity. The $D^{0}$
($\overline{D^{0}}$) $v_{1}$-slope ($dv_{1}/dy$) is calculated by
fitting $v_1(y)$ with a linear function constrained to pass through
the origin, as shown by the solid (dot-dashed) line in
Fig.~\ref{fig2}. The $dv_{1}/dy$ for $D^{0}$ and $\overline{D^{0}}$ is $-0.086 \pm
0.025$ (stat.) $\pm$ 0.018 (syst.) and $-$0.075 $\pm$ 0.024 (stat.)
$\pm$ 0.020 (syst.), respectively. Figure~\ref{fig3}(a) presents
$v_{1}(y)$ averaged over $ D^{0}$ and $\overline{D^{0}}$ (denoted
$\langle v_{1} \rangle$) for $p_{\rm T} >$ 1.5~GeV/$c$. The
$dv_{1}/dy$ for the averaged $D^{0}$ mesons using a linear fit is
$-$0.080 $\pm$ 0.017 (stat.) $\pm$ 0.016 (syst.). The {\it p}-value
and $\chi^{2}$/NDF for the linear fit passing through the origin are
0.41 and 2.9/3 respectively. To perform a statistical significance
test for a null hypothesis for the $v_{1}$ of the averaged  $D^{0}$
and $\overline{D^{0}}$, we calculate the $\chi^{2}$ of the measured
$\langle v_{1} \rangle$ values set to a  constant at zero. The
resulting $\chi^{2}$/NDF and {\it p}-value are 14.9/4 and 0.005
respectively, indicating that the data prefer a linear fit with a
non-zero slope. The $D^{0}$ $v_{1}(y)$ results are compared to charged
kaons, shown by open square markers in Fig.~\ref{fig3}(a). The kaon
$v_{1}(y)$ is measured for $p_{\rm T} >$ 0.2~GeV/$c$. Note that the
$\langle p_{\rm T} \rangle$ for kaons is 0.63 $\pm$ 0.04 GeV/$c$ while
that for $D^{0}$ mesons is  2.24 $\pm$  0.02 GeV/$c$ in our
measured $p_{\rm T}$ acceptance for 10--80\% Au+Au collisions at
$\sqrt{s_{\rm NN}}$ = 200\,GeV. The $dv_{1}/dy$ of charged kaons, fit
using a similar linear function, is $-$0.0030 $\pm$ 0.0001 (stat.)
$\pm$ 0.0002 (syst.). The inset in Fig.~\ref{fig3}(a) presents the
ratio of the $v_{1}$ of the $D^{0}$  and charged kaons. The absolute
value of the $D^{0}$-mesons $dv_{1}/dy$ is observed to be about 25
times larger than that of the kaons with a 3.4$\sigma$
significance. Moreover, among the measurements by the STAR
collaboration of $v_{1}(y)$ for eleven particle species in Au+Au
collisions at 200 GeV~\cite{STAR-BESv1}, the nominal value of the $D^{0}$ $dv_{1}/dy$ is the largest.  

In hydrodynamic models, the ``antiflow'' nature of rapidity-odd directed flow is reproduced by an initial tilted source \cite{tilt_chv1_bozek}, where the tilt parameter is obtained from a fit to $v_{1} (y)$ for charged hadrons. A recent model calculation~\cite{chatt_bozek_dv1_prl}, where Langevin dynamics for heavy quarks are combined with a hydrodynamic medium and a tilted initial source, predicted a larger $v_{1}$ slope for $D$ mesons compared to light hadrons. It has been argued that the large $dv_1/dy$ for $D$ mesons is driven by the drag from the tilted initial bulk medium. A noteworthy feature in Ref.~\cite{chatt_bozek_dv1_prl} is the sensitivity of $dv_1/dy$ for $D$ mesons to the tilt parameter. Ref.~\cite{chatt_bozek_dv1_prl} predicts that the $dv_1/dy$ for $D$ mesons can be 5$-$20 times larger than for charged hadrons, in qualitative agreement with our data,  depending on the choice of tilt and drag parameters. 

An initial transient EM field can induce an opposite $v_{1}$ for charm
and anti-charm quarks. The magnitude of such an induced $v_{1}$ is
predicted to be several orders of magnitude larger than that for light
hadron species due to the early formation of charm
quarks~\cite{das_greco_dv1_em, gursoy_em}. Recently, the authors of
Ref.~\cite{chatt_bozek_dv1_prl} updated their model calculations, and
predicted that the $D$-meson $v_1$ contribution from the tilted
initial source dominates over the contribution from the initial
EM-field~\cite{chatt_bozek_dv1_2nd}. The measured $D^0$ $\langle
v_1(y) \rangle$ is compared to such model calculations (solid line) in
Fig.~\ref{fig3}(a). The model comparison for $D^0$ plus $\overline{D^0}$ indicates that the model gives the correct sign of $dv_1/dy$ but the $v_1$ magnitude is underestimated when using the model parameters of Ref.~\cite{chatt_bozek_dv1_2nd}. The current measurements could help to constrain the model parameters such as the tilt and charm drag coefficients. 

In Fig.~\ref{fig3}(a), the $\langle v_{1} \rangle$ measurements are also compared to a calculation using A-Multi-Phase-Transport  (AMPT) model~\cite{ampt_dv1} shown by the dashed line. In this calculation, although the initial rapidity-odd eccentricity (in spatial coordinates) for heavy quarks is smaller than for light quarks, the magnitude of $v_{1}$ for heavy flavor hadrons is approximately seven times larger than that for light hadrons at large rapidity. The AMPT calculation also suggests that, as a result of being heavy and produced early, the charm hadrons have an enhanced sensitivity to the initial dynamics, over that for light hadrons. This calculation underpredicts the data.

Figure~\ref{fig3}(b) shows the difference between $D^{0}$ and $\overline{D^{0}}$ $v_{1} (y)$ (denoted $\Delta v_{1}$) measured in 10--80\% centrality Au+Au collisions at $\sqrt{s_{\rm NN}}$ = 200\,GeV.  The $\Delta v_{1}$ slope is fitted with a linear function through the origin to give $-0.011 \pm 0.034$ (stat.) $\pm$ 0.020 (syst.).  The dashed and solid lines in Fig.~\ref{fig3}(b) presents the $\Delta v_{1}$ expectation from two models. The solid line (labeled "Hydro+EM") is the expectation from the model with effects from both a tilted source and an initial EM field~\cite{chatt_bozek_dv1_2nd}, while the dotted line is the expectation from the initial EM field only~\cite{das_greco_dv1_em}. From these models, the predicted $\Delta v_{1}$ slope for the charm hadrons lie within the range -0.008 to -0.004. However, different values of medium conductivity and time evolution of the EM fields, as well as the description of charm quark dynamics in the QGP can cause large variations in the charge dependent $v_{1}$ splitting. The present predictions of $\Delta v_{1}$ are smaller than the current precision of the measurement. Nonetheless, the measurement could provide constraints on the possible variations of the parameters characterizing the EM field and charm quark evolution in the QGP.

In summary, we report the first observation of rapidity-odd directed flow ($v_1(y)$) for $D^0$ and $\overline{D^0}$ mesons separately, and for their average, in 10--80\% central Au+Au collisions at $\sqrt{s_{\rm NN}}$ = 200 GeV using the STAR detector at RHIC. The $v_1$ slope ($dv_1/dy$) of $D^0$ mesons are observed to be about a factor of 25 times larger than that for charged kaons with a 3.4$\sigma$ significance. The observation of a relatively larger and negative $v_1$ slope for charmed hadrons with respect to the light flavor hadrons can be qualitatively explained by a hydrodynamic model with an initially tilted QGP source~\cite{chatt_bozek_dv1_prl} and by an AMPT model calculation. These data not only give unique insight into the initial tilt of the produced matter, they are expected to provide improved  constraints for the geometric and transport parameters of the hot QCD medium created in relativistic heavy-ion collisions.

We thank the RHIC Operations Group and RCF at BNL, the NERSC Center at LBNL, and the Open Science Grid consortium for providing resources and support.  This work was supported in part by the Office of Nuclear Physics within the U.S. DOE Office of Science, the U.S. National Science Foundation, the Ministry of Education and Science of the Russian Federation, National Natural Science Foundation of China, Chinese Academy of Science, the Ministry of Science and Technology of China and the Chinese Ministry of Education, the National Research Foundation of Korea, Czech Science Foundation and Ministry of Education, Youth and Sports of the Czech Republic, Hungarian National Research, Development and Innovation Office (FK-123824), New National Excellency Programme of the Hungarian Ministry of Human Capacities (UNKP-18-4), Department of Atomic Energy and Department of Science and Technology of the Government of India, the National Science Centre of Poland, the Ministry  of Science, Education and Sports of the Republic of Croatia, RosAtom of Russia and German Bundesministerium fur Bildung, Wissenschaft, Forschung and Technologie (BMBF) and the Helmholtz Association.


\bibliography{reference.bib}

\end{document}